\documentclass{stsci_report}

\usepackage{graphicx}
\usepackage{siunitx}
\usepackage{todonotes}
\usepackage{pdfpages}
\usepackage{hyperref}

\copyrighttext{Copyright\copyright\ \the\year\ The Association of Universities for Research in Astronomy, Inc. All Rights Reserved.}

\presubtitle{OTE Science Performance Memo JWST-STScI-008497}

\title{ \textbf{\LARGE A Year of Wavefront Sensing with JWST in Flight:  Cycle 1 Telescope Monitoring \& Maintenance Summary}}
\author{Charles-Philippe Lajoie, Matthew Lallo, Marcio Mel\'endez, Nicolas Flagey, Randal Telfer, Thomas M. Comeau, Bernard A. Kulp, Tracy Beck, Gregory R. Brady, and
Marshall D. Perrin*\\
\vspace{0.1cm}
\textit{\footnotesize *Corresponding author. mperrin@stsci.edu}
 }

\date{\today}

\bibliography{references}{}

\begin{document}

\maketitle

%\footnote{Corresponding author: mperrin@stsci.edu}

\abstract{\noindent We summarize JWST's measured telescope performance across science Cycle 1. The stability of the segments' alignment is typically better than 10 nanometers RMS in wavefront error between measurements taken two days apart, leading to highly stable point spread functions. The frequency of segment ``tilt events'' has decreased significantly, and larger tilt events ceased entirely, as structures gradually equilibrated after cooldown. Mirror corrections every 1--2 months now maintain the telescope below 70 nm RMS wavefront error. Observed micrometeoroid impacts during cycle 1 had negligible effect on science performance, consistent with preflight predictions. As JWST begins Cycle 2, its optical performance and stability are equal to, and in some ways better than, the performance reported at the end of commissioning. 
}

\section{Introduction} \label{sec:intro}

%The alignments of the 18 mirrors which comprise JWST's 6.5 m segmented primary are actively sensed and controlled, initially for alignment after launch, and on an ongoing basis to maintain optical performance throughout the science mission. 
Following JWST's launch, the segmented primary mirror was aligned to diffraction-limited precision using focus-diverse wavefront sensing and control \parencite{Acton_2022SPIE12180E..0UA, Feinberg+Wolf+Acton+etal_2022}. 
Now in science operations, the telescope alignment has continued to be measured roughly every 48 hours, and mirror corrections have been commanded as needed to maintain optical performance. We report here on the results of the Cycle 1 wavefront sensing (WFS) program and describe the measured stability of JWST after the first year of science operations. For further details on the JWST Optical Telescope Element (OTE) and WFS methods, see \cite{McElwain2023PASP..135e8001M}.

%That requirement is achieved with margin; during cycle 1,  the average time between correction is 3.8 weeks, and
Many factors affect the stability of large space optical systems, including but not limited to changes in spacecraft thermal state, vibrations and dynamical excitations, and space environmental effects such as micrometeoroid impacts and space weathering. Another source of wavefront variation is ``tilt events'', i.e. occasional abrupt shifts in mirror positions thought to be caused by stick-slip release of stored stresses from thermal contraction during cooldown. As discussed further below, these events have gradually decreased in frequency as structures equilibrate over time. JWST was designed for high stability, including requirements to need mirror corrections no more often than every two weeks, and to maintain point spread function (PSF) encircled energy stability better than 2.5\% over 14 days even after thermally worst-case slews. Achieving this level of stability required substantial efforts in observatory design, systems engineering, and verification \parencite{Menzel2023PASP..135e8002M}. Commissioning results confirmed these requirements were met and/or surpassed, and allowed quantifying stability on multiple timescales \parencite{McElwain2023PASP..135e8001M, Rigby2023PASP..135d8001R}. 

WFS observations are scheduled roughly every two days, using NIRCam to obtain defocused ($\pm8$ waves) images of bright stars at 2.12 $\mu$m. Targets are drawn from an all-sky pool so as to minimize slew overheads between science and WFS. Each WFS measurement includes any actual alignment changes plus some sensing noise, which is $\sim5-7$~nm RMS for JWST. The known NIRCam instrument wavefront contribution is subtracted to yield a measured wavefront for the telescope alone. Each measurement is fit into piston, tip, and tilt (henceforth ``PTT'') optical modes per segment. Those PTT values and overall RMS wavefront error (WFE) are compared against thresholds to determine when mirror corrections are necessary.
%\footnote{JWST cannot adjust its mirrors while taking science observations, so less frequent mirror corrections maximizes time for science.}
All WFS data and derived wavefront maps are available in the MAST archive\footnote{MAST DOI: \href{http://dx.doi.org/10.17909/dyck-vk86}{10.17909//dyck-vk86} and \href{https://outerspace.stsci.edu/display/MASTDOCS/Data+Search\#DataSearch-SearchforWavefrontProducts}{``JWST WSS'' Collection}.} and may be retrieved using, for instance, the \texttt{webbpsf} package \parencite{Perrin+Sivaramakrishnan+Lajoie+etal_2014}. 

% https://outerspace.stsci.edu/display/MASTDOCS/Supplemental+Products#SupplementalProducts-WavefrontSensingData
 
%\todo{Recap key facts of WFSC. Roughly 7~nm RMS sensing noise. The actuator step size of $\sim10$~nm limits the precision of control. When needed, corrections are applied on a subsequent WFS visit. Correction threshhold was initially set at WFE >80~nm, but given excellent performance partway through cycle 1 that was revised to 70~nm.}
%\todo{Want to minimize control due to time cost of control}

%\clearpage

\section{Results} \label{sec:results}

\subsection{Observed wavefront stability}

\begin{figure*}[!p]
    \centering
    \includegraphics[width=\textwidth]{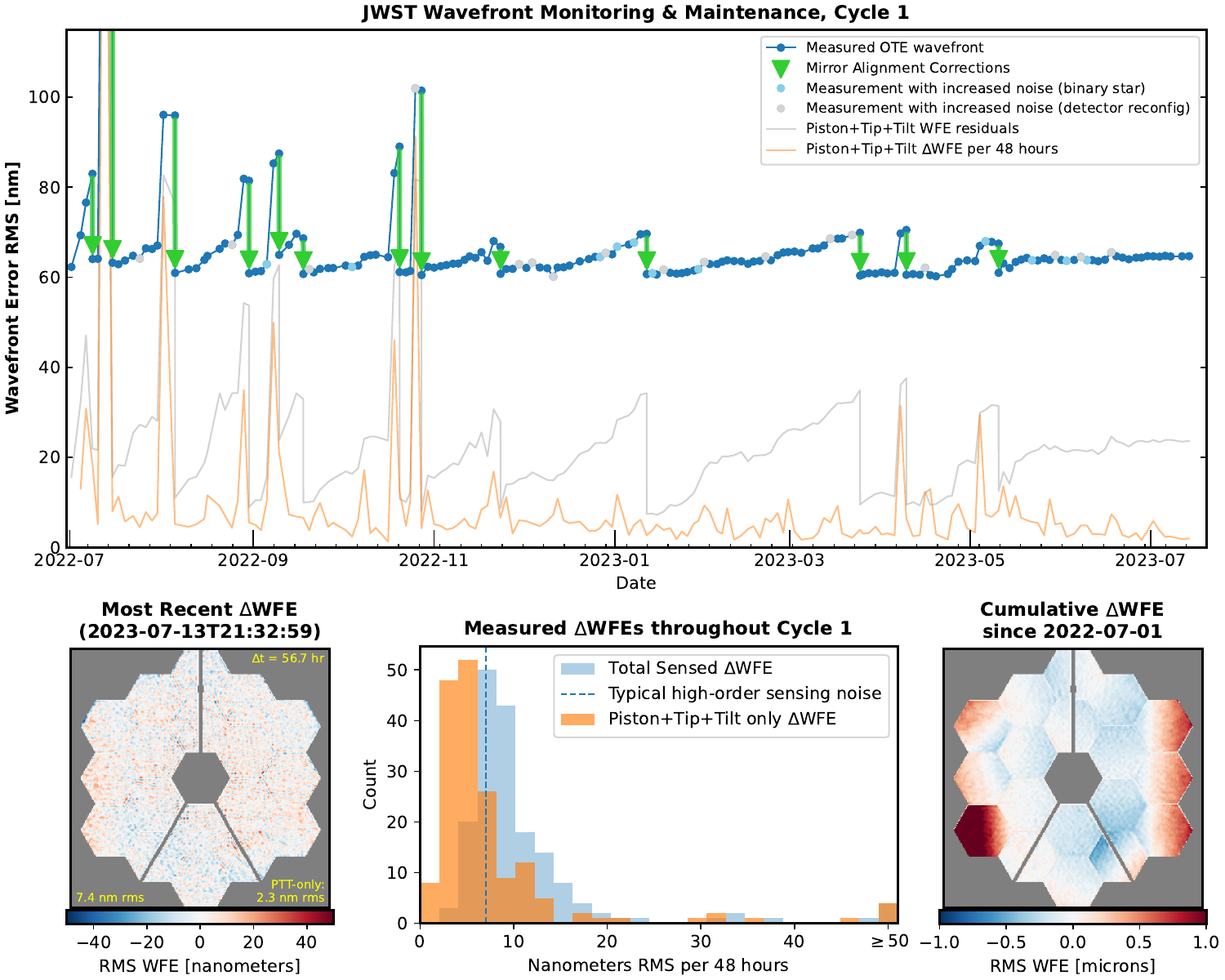}
    \caption{Summary of Cycle 1 JWST WFS results, including dates up to 2023 July 13. \textit{Top panel:} WFS measurements and corrections over time. Individual WFS measurement results are shown in blue, with green arrows indicating the occasional mirror corrections. The measurements are fit to derive the correctable segment PTT modes at any given time (gray) and the change in PTT modes per two day measurement period (orange). 
    \textit{Lower left:} Measured stability from the most recent WFS as of this writing. Changes are almost unmeasurable for most segments, with a total of only 2.3~nm RMS in PTT modes. This stability is typical of recent measurements.  \textit{Lower middle:} Histogram of stability levels over one year. The observed median stability per 48 hours is 9.0~nm RMS in total measured WFE (including measurement noise), and only 5.3~nm RMS in PTT modes. \textit{Lower right:} Cumulative drift corrected by mirror moves to maintain alignments. The relaxation of the ``wing'' segments, outboard of the backplane hinge lines, has been the dominant effect. 
    \label{fig:one}
    }
\end{figure*}

Figure 1 summarizes the results of telescope monitoring and control throughout cycle 1. Three distinct regimes can be discerned:  Early on, there were a handful of abrupt step-function jumps in WFE, the results of larger tilt events. In between and after those events, the WFE has generally increased slowly, the result of gradual accumulation of smaller tilt events and other mirror drifts. Finally, most recently, there have been periods of time during which WFE has remained flat, with near-zero drifts over time.  

%The telescope WFE was maintained below the 80~nm RMS control threshold for 94\% of this time. 
A key result from cycle 1 is the observed increase in telescope stability over time. The last observed large ($>50$~nm RMS segment-level WFE) tilt event occurred in 2022 October. Small ($<10$~nm RMS) tilt events continue at reduced rates, and there have been only two moderate events ($\sim30$~nm RMS) since 2022 October. The most recent period starting mid-May 2023 has been the most stable yet. 

The changes in WFE (henceforth, $\Delta$WFE) between successive measurements taken $\sim$48 hours apart are very small, typically $<10$~nm RMS including the measurement noise. 
%We normalized the measured $\Delta$WFEs to account for the scatter in elapsed times between WFS measurements, scaling to units of nanometers per 48 hours.  
The median $\Delta$WFE during cycle 1 was 9.0 nm RMS per 48 hours for the total measured WFE (see Figure 1 lower middle). Considering just the PTT modes, the median $\Delta$WFE is only 5.3 nm RMS per 48 hours.\footnote{Equivalently, 110 picometers per hour, though JWST does not in practice achieve on short timescales that stability level. In particular, the cycling of heaters in the Instrument Electronic Compartment drives $\sim 2$~nm oscillations of mostly astigmatism on 3--4 minute timescales. The regular WFS monitoring on longer timescales mostly averages over that. } %Half of measurements have PTT stability levels below that, approaching unmeasurable even with JWST's very precise WFS.
No significant dependence on observatory attitudes or slew history has been observed; efforts to better characterize and model the telescope behavior are ongoing as additional data become available.

The stable wavefronts lead to similarly stable PSFs: the encircled energy stability is required to be $<2.5\%$ at 0.08 arcsec, but is generally $<0.5\%$ and, in recent months is much $\ll0.5\%$. 

The decrease in tilt events over time provided an opportunity to tighten the control threshold for mirror realignments, initially set to 80~nm WFE but now 70~nm since mid-cycle-1. Since 2022 November 1, the telescope wavefront has been maintained below 70~nm for $\sim85\%$ of the time. Only twelve mirror corrections were needed during the cycle, mostly in the first four months. More recently, corrections have typically been one to two months apart.

The cumulative sum of all mirror drifts observed in cycle~1 (Figure 1 lower right) is dominated by the side ``wing'' segments, which show roughly similar patterns of positive $\Delta$WFE outward. This is particularly noteworthy for the right (``+V2") wing, for which distinct tilt events of individual segments gradually summed into a strikingly uniform tilt of the entire wing. This behavior is consistent with the hypothesis that the driver of tilt events is differential thermal contraction stresses, particularly from dissimilar materials along the hinge line which was actuated and latched at $\sim 100$~K and subsequently cooled to 35~K. 
Thus, the tilt event phenomenon would likely be less significant for a deployable but non-cryogenic observatory.

%\subsection{Measured Mirror Stability and Frequency of Control}

%\todo{Excellent stability, median PTT delta is 5.3~nm RMS.  Median controllable PTT residual is 21~nm}

\subsection{Other Factors}

Some WFS measurements have slightly increased noise: A few target stars proved to have previously-unknown stellar companions (such stars are then removed from the target pool for future measurements).  Intermittently, anomalous detector behavior has been observed after reconfiguring from the special $8\times8$ subarray used for the jitter measurements. This effect, a rare detector electronics anomaly known pre-launch, results in a ``bright channel'' effect with excess counts in particular columns and an increase in crosstalk\footnote{This effect has not been observed with any other subarray, and is not known to affect any science data.}. These effects are sufficiently small to not significantly impact wavefront monitoring, but at times lead to slight apparent outliers from the overall trend. Affected measurements are indicated with distinct colors in Figure 1 and are included for completeness. 

Additionally, separate analyses assessed, but did not find, wavefront focus variations that would indicate secondary mirror drifts. The secondary mirror and its support structure are stable to better than the sensing precision. 

Each WFS observation also includes a measurement of line-of-sight pointing jitter (see \cite{Hartig+Lallo_2022..JWST-STScI-008271}), which is consistently below 1 milliarcsecond. No dependence on reaction wheel speeds or stored momentum has been observed.

%\todo{Instrument heater oscillations}

Micrometeoroid impacts are detected 
%as few-nanometer localized changes in individual mirrors 
periodically in WFS data as well as in  pupil imaging observations (collected four times a year). The impact rates have been consistent with preflight predictions and no additional large impacts were detected, which is consistent with the 2023 May impact being an outlier. The several dozen minor impacts during cycle 1, cumulatively contributing $<1$~nm RMS increase in global WFE, were all sufficiently small to have no measurable effect on science performance. We note that the radius-of-curvature actuator on each segment can be used on occasion to mitigate the small cumulative effects of micrometeoroid impacts over the mission lifetime. Moreover, the Micrometeoroid Avoidance Zone policies adopted for cycle 2 should reduce impact rates long-term. 

\section{Future Work}

The WFS monitoring program will continue unchanged into Cycle 2. Ongoing efforts will further characterize and model JWST's stability and optical performance over time. 

A small handful of specialized datasets provide measurements of JWST wavefronts on timescales shorter than the 48 hour monitoring cadence. These include several engineering datasets obtained during commissioning as well as science observations obtained using the NIRCam Grism Time Series Observations mode, which use a defocus weak lens to obtain time-series photometry in NIRCam's short wave channel (see \cite{Schlawin2023PASP..135a8001S}). Ongoing analyses of these data are yielding finer and faster views of JWST wavefront stability, which will be reported in a future work (Telfer et al. in prep). These data confirm that, after subtracting out the short-timescale wavefront oscillations driven by instrument heaters, on intermediate timescales the telescope structure does, at least in some cases, achieve the $\sim$ 0.1 nanometers per hour stability level inferred based on the 48-hour sensing results. 

JWST's superb stability demonstrates how effectively the mission's development and integrated modeling and testing achieved highly stable structures. This provides a strong technical heritage toward future missions requiring even greater ultra-stability.

\section*{Acknowledgements}

%\begin{acknowledgments}
This report was made possible by the many individuals from NASA, Ball Aerospace, Northrop Grumman, and STScI who contributed to JWST over the years.

\vspace{5mm}

\noindent This work is based on observations made with the NASA/ESA/CSA James Webb Space Telescope. The data were obtained from the Mikulski Archive for Space Telescopes at the Space Telescope Science Institute, which is operated by the Association of Universities for Research in Astronomy, Inc., under NASA contract NAS 5-03127 for JWST. These observations are associated with calibration programs 2586, 2724, 2726, and 2727.
%, and their partner organizations and subcontractors.
%\end{acknowledgments}

\vspace{5mm}
%\facilities{JWST}


\noindent Software used:\\
\texttt{webbpsf} \parencite{Perrin+Sivaramakrishnan+Lajoie+etal_2014}, \\
jwst pipeline \parencite{jwst_data_pipeline},\\
JWST Wavefront Sensing Subsystem \parencite{Perrin_etal_2016_SPIE}
 


\printbibliography

%\include{references.bbl}



\end{document}